\documentclass[a4paper,11pt]{article}
\usepackage[super,compress]{cite}
\usepackage{authblk}
\usepackage{graphicx}
\usepackage{amsmath}
\usepackage{amssymb}
\usepackage{hyperref}
\usepackage{appendix}
\usepackage{graphicx}
\usepackage{xcolor}

\title{A 331 Model for Lepton Flavor Universality Violation in \(B\) Decays}
\author[1,3]{Soumia Lebbal\thanks{\href{mailto:soumia.lebbal@univ-batna.dz}{soumia.lebbal@univ-batna.dz}}}
\author[2]{Jamal Mimouni\thanks {\href{mailto:jamalmimouni@umc.edu.dz}{jamalmimouni@umc.edu.dz}}}
\author[3]{Noureddine Mebarki\thanks{\href{mailto:nnmebarki@yahoo.fr}{nnmebarki@yahoo.fr}}}
\affil[1]{\small{Laboratoire de Physique des Rayonnements et de leurs Interactions avec la Matière (PRIMALAB), Physics Department, University of Batna 1, Allées 19 Mai, Route de Biskra - Batna, 05000 Algeria}}
\affil[2]{\small{Laboratoire de Physique Mathématique et Physique Subatomique (LPMPS), Physics Department, University Frères Mentouri Constantine 1, Ain El-Bey Road, Constantine, Algeria, 25017 \& CERIST}}
\affil[3]{\small{Laboratoire de Physique Mathématique et Physique Subatomique (LPMPS), Physics Department, University Frères Mentouri Constantine 1, Ain El-Bey Road, Constantine, Algeria, 25017}}

\begin{document}
\date{}
\maketitle

\begin{abstract}
Lepton Flavor Universality Violation (LFUV) in \(B\) decays in both neutral \(b\rightarrow s l^+ l^-\) and charged \(b\rightarrow cl\bar{\nu}_{l}\) processes observed in the ratios \(R_{K^{(*)}}\) and \(R_{D^{(*)}}\) is investigated in the framework of a model based on the extended symmetry gauge \(SU(3)_{C}\otimes SU(3)_{L}\otimes U(1)_{X}\) whose leptonic sector consists of five triplets. It is shown that in order to explain the experimentally observed deviations from the Standard Model in these flavor changing transitions, a non-minimal version of the model has to be considered. We investigate the ability of this model in accommodating the model-independent scenarios currently favored by global fits.

\end{abstract}

\section{Introduction}	

Experimental hints of Lepton Flavor Universality Violation (LFUV) have appeared in semi-leptonic B-decays, both in charged and in neutral processes. In fact, disagreements with the SM expectations have shown contributions of non-SM origin of size \(\mathcal{O}(10\%)\) compared to the corresponding SM amplitudes. In particular, four anomalies have appeared in ratios assessing LFU in the decays \(B\longrightarrow D^{(*)}l\bar{\nu}\), \( (l=e,\mu)\) and \(B\longrightarrow K^{(*)}l^{+}l^{-}\). According to the underlying quark transition, the ratios of interest reported by LHCb, can be grouped into two categories: ones that assess deviations from \(\mu / \tau\) and \(\tau / e\) universality in  \(b\rightarrow cl\bar{\nu}_{l}\) charged currents \cite{LHCb 2015}, and the ones that assess deviations from \(\mu / e\) universality in \(b\rightarrow s l^+ l^-\) neutral currents \cite{LHCb 2017, LHCb 2019}. They are defined, respectively, as 
\[R_{D^{(*)}}=\frac{\mathcal{B}\left(B\longrightarrow D^{(*)}\tau\bar{\nu}_{\tau} \right)}{\mathcal{B}\left(B\longrightarrow D^{(*)} l\bar{\nu}_{l}  \right)}, [l=e, \mu],\]
\[R_{K^{(*)}}=\frac{\mathcal{B}\left(B\longrightarrow K\mu^{+}\mu^{-} \right)_{q^{2}\in [q^{2}_{min}, q^{2}_{max}]}}{\mathcal{B}\left(B\longrightarrow K e^{+} e^{-} \right)_{q^{2}\in [q^{2}_{min}, q^{2}_{max}]}},\]
where \(R_{K^{(*)}}\) are measured over specific ranges for the squared dilepton invariant mass \(q^{2}\) (in \(GeV^{2}/c^{4}\)). Furthermore, LHCb has reported a strong evidence for a deviation from the SM observed in the angular distribution of the \(B^0\longrightarrow K^{*0}\mu^{+}\mu^{-}\) which is consistent with the LFUV in \(B\) decays \cite{angular distribution}. A series of theoretical speculations about a possible New Physics (NP) interpretations have emerged. Within a model independent approach, many  analyses have been performed in terms of Effective Field Theories (EFTs){ corresponding to the SM at the \(b\)-quark mass supplemented with additional lowest dimensional non-SM operators \cite{angular distribution, model independant approach1, model independant approach2}. Other approaches are model dependent which require choosing a specific scenario of physics beyond the SM, and try to see if this model allows for such anomalies. A class of particularly motivated models includes those which are based on new (heavy) exotic bosonic mediators at the TeV scale that couple to the lepton and the quark generations differently \cite{model1, model2, model3}. A good candidate of such models is the 331 model, whose heavy (exotic) gauge bosons are LFUV processes mediators. In order to generate LFUV processes, the coupling of the model's gauge bosons with the fermions have to differ between generations. This requires the extension of the leptonic sector by introducing two additional generations. A previous study of such non-minimal version with \(\beta=-1/\sqrt{3}\) \cite{main article} has succeeded in explaining LFUV in the Flavor Changing Neutral Current transition (FCNC) \(b\rightarrow s\). It has been shown that, under the assumption that LFUV is dominated by neutral gauge boson contributions, the model could reproduce scenarios with NP contributions in (\(C_9^{\mu},C_{10}^{\mu}\)) in good agreement with global fit analyses of \(b\rightarrow sll\). In the present work, we attempt to do the same for a non-minimal version of the 331 model with \(\beta=1/\sqrt{3}\) for both, Flavor Changing Charged Current transitions (FCCC) \(b\rightarrow c\) and the FCNC transition \(b\rightarrow s\).     \\
The paper is organized as follows: in Sec. 2 we review the main features of the non-minimal 331 model for our specific choice of the parameter \(\beta\). In Sec. 3, we describe the general framework of our study for both transitions: FCCC and FCNC by reviewing \(B\) decay processes within the EFT approach. Sec. 4, is dedicated to the model's gauge boson-mediated NP contribution for both types of processes. In Sec. 5, we present these contributions and compare them to the global analysis scenarios that are currently favored by the data, and finally, in Sec. 6,  we draw our conclusion.

\section{The Model}
We extend the SM by enlarging its gauge group to the broader \(SU(3)_{C}\times SU(3)_{L}\times U(1)_{X}\) group. The minimal construction \cite{minimal construction} is based on placing the left-handed lepton doublets in \(SU(3)_{L}\) triplets that transform in the same way, while the flipped set-up \cite{flipped set-up} is based on perfect quark family replication. In both cases, the gauge bosons couple identically to all lepton and quark families, respectively. Thus, no LFUV can arise from these couplings. In order to generate LFUV from different couplings of the gauge bosons to all fermionic fields, we adopt the non-minimal 331 construction (Model B in Ref. \citen{anomaly-free set}). In this construction, the leptons should be grouped in no less than 5 generations (appendix C in Ref. \citen{main article}). The electric charge generator in \(SU(3)_{L}\times U(1)_{X}\) is given by
\begin{equation}
Q=T_{3L}+\beta T_{8L}+XI_{3},
\end{equation}
where \(T_{8L}=(1/2\sqrt{3})\)diag\((1,1,-2)\) and \(T_{3L}=(1/2)\)diag\((1,-1,0)\) are the \(SU(3)_{L}\) diagonal generators, \(X\) is the quantum number associated with \(U(1)_{X}\) and \(I_{3}=\) diag\(\left(1, 1, 1 \right)\). The proportionality constant \(\beta\) distinguishes different 331 models. It is shown that it can have 4 possible values: \(\pm \sqrt{3}, \pm (1/\sqrt{3})\) which  play a key role in determining the electric charges of extra particles and only for some of its values, the gauge bosons have integer charges. For instance, \(\beta=\pm (1/\sqrt{3})\) does not introduce exotic electric charges of fermions in order to cancel the anomalies. Furthermore, for theses value, the heavy gauge bosons have an integer electric charge. The non-minimal set that we adopt for \(\beta=1/\sqrt{3}\) presents a spectrum that contains 16 charged particles with masses in agreement with the observations (no light particles apart from the SM ones): 9 quarks (6 light SM's and 3 exotic heavy ones) and 7 charged leptons (3 SM's and 4 exotic), and 7 neutral ones (3 SM's and 4 exotic).
\subsection{Particle Content and Representations}
For \(\beta=1/\sqrt{3}\) (Model B in Ref. \citen{anomaly-free set2}), we introduce the left-handed components of the fields together with the right-handed partners of the charged ones
\begin{enumerate}
\item[(i)] three generations of quarks
\begin{equation}
\begin{aligned}
\label{quark fields}
&Q_{L}^{m}=
\begin{pmatrix}
d_{m}\\-u_{m}\\U_{m} 
\end{pmatrix}
; u_{m}^{R}; d_{m}^{R}; U_{m}^{R}, &\left( m=1,2\right),
\end{aligned}
\end{equation}
with \(SU(3)_{C}\times SU(3)_{L}\times U(1)_{X}\) quantum numbers \(\left( 3, \bar{3}, \frac{1}{3} \right), \left( \bar{3}, 1, 2/3\right), \left( \bar{3},1,-1/3\right), \left( \bar{3}, 1, 2/3\right)\).
\begin{equation}
\begin{aligned}
\label{quark fields}
&Q_{L}^{3}=
\begin{pmatrix}
u_{3}\\d_{3}\\D_{3} 
\end{pmatrix}
; u_{3}^{R}; d_{3}^{R}; D_{3}^{R},
\end{aligned}
\end{equation}
with \(SU(3)_{C}\times SU(3)_{L}\times U(1)_{X}\) quantum numbers \(\left( 3, 3, 0 \right), \left( \bar{3}, 1, 2/3\right), \left( \bar{3},1,-1/3\right), \left( \bar{3}, 1, -1/3\right)\).
\item[(ii)] five generations of leptons
\begin{equation}
\begin{aligned}
\label{lepton fields}
l_{1}^{L}=
\begin{pmatrix}
e^{-L}_{1}\\-\nu^{L}_{1}\\N^{L}_{1}
\end{pmatrix}; e^{-R}_{1},
\end{aligned}
\end{equation}
with \(SU(3)_{C}\times SU(3)_{L}\times U(1)_{X}\) quantum numbers \(\left( 1, \bar{3}, -1/3 \right), \left( 1, 1, -1\right)\).
\begin{equation}
\begin{aligned}
\label{lepton fields}
l_{n}^{L}=
\begin{pmatrix}
\nu^{L}_{n}\\e^{-L}_{n}\\E^{-L}_{n}
\end{pmatrix}
;&e_{n}^{-R}; E_{n}^{-R}, &\left( n=2,3\right),
\end{aligned}
\end{equation}
with \(SU(3)_{C}\times SU(3)_{L}\times U(1)_{X}\) quantum numbers \(\left( 1, 3, -2/3 \right), \left( 1, 1, -1\right), \left( 1, 1, -1\right)\).
\begin{equation}
\begin{aligned}
\label{lepton fields}
l_{4}^{L}=
\begin{pmatrix}
N^{L}_{4}\\E^{-L}_{4}\\F^{-L}_{4}
\end{pmatrix}
;&F_{4}^{-R},
\end{aligned}
\end{equation}
with \(SU(3)_{C}\times SU(3)_{L}\times U(1)_{X}\) quantum numbers \(\left( 1, 3, -2/3 \right), \left( 1, 1, -1\right)\).
\begin{equation}
\begin{aligned}
\label{lepton fields}
l_{5}^{L}=
\begin{pmatrix}
\left(E^{-R}_{4}\right) ^{c}\\N_5^L\\P^{L}_{5}
\end{pmatrix},
\end{aligned}
\end{equation}
with \(SU(3)_{C}\times SU(3)_{L}\times U(1)_{X}\) quantum numbers \(\left( 1, 3, 1/3 \right)\).
\end{enumerate}
It should be stressed that, originally, the model contains 15 leptons instead of 14 (eight charged and seven neutral ones), but because we have to prevent the existence of exotic particles at the EW scale, we had to identify the fifth generation positively charged lepton with the charge conjugate of the right handed partner of the fourth generation (see \ref{appA}).
The \(SU(3)_{L}\) gauge bosons are denoted in the matrix form by \(W_{\mu}=W_{\mu}^{a}T^{a}\), where \(T^{a}=\lambda^{a}/2\) are the generators of \(SU(3)_{L}\) (\(\lambda^{a}\) being the Gell-Mann matrices and \(a=1,..8)\). For our specific choice of \(\beta\) \cite{gauge bosons}
\begin{equation}
\begin{aligned}
& W_{\mu}=W_{\mu}^{a}T^{a}=\dfrac{1}{2}& 
\begin{pmatrix}
W_{\mu}^{3}+\dfrac{1}{\sqrt{3}}W_{\mu}^{8}& \sqrt{2}W_{\mu}^{+}& \sqrt{2}Y_{\mu}^{+}\\\sqrt{2}W_{\mu}^{-}& -W_{\mu}^{3}+\dfrac{1}{\sqrt{3}}W_{\mu}^{8}& \sqrt{2}V_{\mu}^{+0}\\ \sqrt{2}Y_{\mu}^{-}& \sqrt{2}V_{\mu}^{-0}&-\dfrac{2}{\sqrt{3}}W_{\mu}^{8}
\end{pmatrix},
\end{aligned}
\end{equation}
where
\begin{equation}
\begin{aligned}
&W_{\mu}^{\pm}=\frac{1}{\sqrt{2}}(W_{\mu}^1\mp iW_{\mu}^2),\\
&Y_{\mu}^{\pm}=\frac{1}{\sqrt{2}}(W_{\mu}^4\mp iW_{\mu}^5),\\
&V_{\mu}^{\pm0}=\frac{1}{\sqrt{2}}(W_{\mu}^6\mp iW_{\mu}^7).
\end{aligned}
\end{equation}
The model contains the two SM's charged gauge bosons \(W_{\mu}^{\pm}\) and two additional heavy singly charged ones \(Y_{\mu}^{\pm}\), whereas,  the neutral gauge bosons are \(W_{\mu}^{3}\), \(W_{\mu}^{8}\), \(W_{\mu}^{6}\) and \(W_{\mu}^{7}\). In what follows, we introduce the flavor vectors where we gather the fields of the same electric charge

\begin{equation}
\begin{aligned}
&U=\left( u_{1},u_{2},u_{3},U_{1},U_{2} \right)^{T}, &\\
&D=\left( d_{1},d_{2},d_{3},D_{3}\right)^{T}, &\\
&f^{-}_L=\left( e^{-L}_{1},e^{-L}_{2},e^{-L}_{3},E^{-L}_{2},E^{-L}_{3},E^{-L}_{4},F^{-L}_{4}\right)^{T}, &\\
&f^{-}_R=\left( e^{-R}_{1},e^{-R}_{2},e^{-R}_{3},E^{-R}_{2},E^{-R}_{3},E^{-R}_{4},F^{-R}_{4}\right)^{T}, &\\
&N_L=\left( \nu_{1}^{L},\nu_{2}^{L},\nu_{3}^{L},N^{L}_{1},N^{L}_{4},N^{L}_{5},P^{L}_{5}\right)^{T}. &\\
\end{aligned}
\end{equation}
\subsection{Symmetry Breaking and Gauge Bosons' Spectrum}
The model undergoes two stages of Spontaneous Symmetry Breaking (SSB). The first, triggered by a sextet and a triplet \eqref{first SSB}, occurs at an energy scale \(\Lambda_{NP}\sim TeV\) and allows one to recover the SM. At the order of \(\Lambda_{NP}\), all exotic charged particles acquire mass: seven fermions (four leptons and three quarks), and the guage bosons \(W_{\mu}^{4}\), \(W_{\mu}^{5}\), \(W_{\mu}^{6}\) and \(W_{\mu}^{7}\). The two neutral \(X_{\mu}\) and \(W_{\mu}^{8}\) yield a massive \(Z_{\mu}^{'}\) and a massless one \(B_{\mu}\), with a mixing angle \(\theta_{331}\)
\begin{equation}
\begin{pmatrix}
Z_{\mu}^{'}\\B_{\mu}
\end{pmatrix}
=
\begin{pmatrix}
\cos\theta_{331}& \sin\theta_{331}\\-\sin\theta_{331}& \cos\theta_{331}
\end{pmatrix}
\begin{pmatrix}
X_{\mu}\\W^{8}_{\mu}
\end{pmatrix},
\end{equation} 
where 
\begin{equation}
\begin{aligned}
&\sin\theta_{331}=\frac{g}{\sqrt{g^{2}+\frac{g^{2}_{X}}{18}}}, &\cos\theta_{331}=\frac{\frac{g_{X}}{3\sqrt{2}}}{\sqrt{g^{2}+\frac{g^{2}_{X}}{18}}}.&
\end{aligned}
\end{equation}
Here, \(g\) and \(g_{X} \) are the gauge coupling constants. The subsequent SSB occurs at energy scale \(\Lambda_{EW}\). It reproduces the EWSB of the SM and is accomplished by means of two triplets and three sextets \eqref{second SSB}. At this stage, the remaining SM fermions and the three gauge fields should all acquire mass. The neural bosons \(W_{\mu}^{3}\) and \(B_{\mu}\) mix together with a mixing angle \(\theta_{W}\) (Weinberg angle) to yield  a massless \(A_{\mu}\) identified with the photon, and a massive \(Z_{\mu}\) that includes components along the exotic fields \(Z^{'}_{\mu}\) and \(W^6_{\mu}\) \cite{anomaly-free set}. In order to keep track of the magnitude of the model's NP contribution, we introduce the parameter \(\epsilon=\Lambda_{EW}/\Lambda_{NP}\) (\(\Lambda_{EW}\ll \Lambda_{NP}\)). After diagonalizing the neutral gauge bosons mass matrix, we see that these two additional components (\(Z^{'}_{\mu}\) and \(W^6_{\mu}\)) should be neglected due to their \(\epsilon^2\) (or higher) suppression, and because, as we will see, the \(Z_{\mu}\) mediated neutral transition \(b\longrightarrow s\) is already suppressed by \(\epsilon^2\). Thus, the EWSB neutral gauge bosons are 
\begin{equation}
\begin{pmatrix}
Z_{\mu}\\A_{\mu}
\end{pmatrix}
\sim
\begin{pmatrix}
\cos\theta_{W}& -\sin\theta_{W}\\ \sin\theta_{W}& \cos\theta_{W}
\end{pmatrix}
\begin{pmatrix}
W^{3}_{\mu}\\B_{\mu}
\end{pmatrix},
\end{equation} 
where, the two mixing angles \(\theta_{331}\) and \(\theta_{W}\) and the two gauge coupling constants obey the relations
\begin{equation}
\label{relation between two gauge coupling constants}
\begin{aligned}
&\cos\theta_{331}=\frac{1}{\sqrt{3}}\tan\theta_{W},&  \frac{g}{g_{X}}=\frac{\tan\theta_{331}}{3\sqrt{2}}.&
\end{aligned}
\end{equation}
Before proceeding, we should mention that the neutral leptons are left out of the discussion since, as we will see, they do not affect the study for the FCNC transition.
\section{\(B\) decay within the Effective Theory Approach}
Assuming that NP originates at a scale \(\Lambda_{NP}\) far above the EW scale (\(\Lambda_{EW}\)). In the window above the EW scale and below the NP scale, NP effects are described within an Effective Field Theory (EFT) approach where the heavy fields are integrated out and absorbed into short-distance coefficients (Wilson Coefficients) multiplying six-dimensional operators built out of light (SM) fields \cite{D6, D6'}. Weak meson decay processes are described within the framework of EFTs. The effective Hamitonian approach can be applied to both FCCC and FCNC.
\subsection{Charged Current \(b\rightarrow cl\bar{\nu}_{l}\) Transition }
For the quark-level transition (\(b\rightarrow c l^{-}\bar{\nu}_{l^{'}})\), the most general elementary charged-current Hamiltonian mediating semi-leptonic decays reads
\begin{equation}
\label{SM Hamiltonian}
\mathcal{H}_{eff}\left( b\rightarrow c l^{-}\bar{\nu}_{l^{'}}\right)=\mathcal{H}^{SM}_{eff}+\mathcal{H}^{NP}_{eff}  =\frac{4G_{F}}{\sqrt{2}}V_{cb} \left(\mathcal O_{V_{L}}^{ll^{'}} \delta_{ll^{'}} +\sum_{i} C^{ll^{'}}_i\mathcal O_i + h.c\right),
\end{equation}
where the sum runs over all dimension-six operators \(\mathcal{O}_i\) (\(i \in \lbrace V_{L(R)},  S_{L(R)}, T\rbrace\)) allowed by the SM gauge symmetry \cite{Operators, Ops.2}.
\begin{equation}
\label{Operators}
\begin{aligned}
\mathcal{O}_{V_{L}}^{ll^{'}}=\left( \bar{c}\gamma_{\mu} P_L b\right) \left( \bar{l}\gamma^{\mu}U P_L\nu_{l^{'}}\right), &
&\mathcal{O}_{V_{R}}^{ll^{'}}=\left( \bar{c}\gamma_{\mu} P_R b\right) \left( \bar{l}\gamma^{\mu}U P_L\nu_{l^{'}}\right),&\\
\mathcal{O}_{S_{L}}^{ll^{'}}=\left( \bar{c}P_L b\right) \left( \bar{l} U P_L \nu_{l^{'}}\right),&
&\mathcal{O}_{S_{R}}^{ll^{'}}=\left( \bar{c} P_R b\right) \left( \bar{l} U P_L \nu_{l^{'}}\right),\\
\mathcal{O}_{T}^{ll^{'}}=\left( \bar{c}\sigma^{\mu\nu} P_L b\right) \left( \bar{l}\sigma_{\mu\nu}U P_L \nu_{l^{'}}\right).
\end{aligned}
\end{equation}
\(V_{cb}\) is the CKM matrix element and \(U\) stands for the PMNS matrix. The Wilson Coefficients \(C_i\) quantify the deviations from the SM. The coefficients \(C_{V_{L}}^{ll^{'}}\) are defined such that they vanish in the SM. In the following, we will be interested in only the \(V-A\) form solution of the NP \cite{fav.solution} that can only come from the the model's charged gauge bosons \(W_{\mu}^{\pm}\) and  \(Y_{\mu}^{\pm}\). Thus, we will not be interested in the coefficients \(C_S\) and \(C_T\) for the scalar and tensor operators. Taking the SM neutrino, and for the same lepton flavor transition (\(l \equiv l^{'}\)), the only significant operator is the product of the two left-handed currents \(\left( \bar{c}_{L}\gamma_{\mu}b_{L}\right) \left( \bar{l}_{L}\gamma^{\mu}\nu_{L}\right)\)
\begin{equation}
\label{total Hamiltonian}
\begin{aligned}
\mathcal{H}^{SM}_{eff}\left( b\rightarrow c l^{-}\bar{\nu}_l\right)=\frac{4G_{F}}{\sqrt{2}}}V_{cb} \left( \bar{c}_{L}\gamma_{\mu}b_{L}\right) \left( \bar{l}_{L}\gamma^{\mu}\nu_{L}\right)+h.c. 
{\end{aligned}
\end{equation}

\begin{equation}
\begin{aligned}
\label{VA contribution}
\mathcal{H}^{NP}_{eff}\left( b\rightarrow c l^{-}\bar{\nu}_l\right)= \frac{4G_{F}}{\sqrt{2}}V_{cb}C^{ll}_{V_{L}} \left( \bar{c}_{L}\gamma_{\mu}b_{L}\right) \left( \bar{l}_{L}\gamma^{\mu}\nu_{L}\right)+h.c.  
\end{aligned}
\end{equation}
\(C^{ll}_{V_{L}}\equiv C_{V_{L}}\) is the Wilson Coefficient of interest.
\subsection{Neutral Current \(b\rightarrow s l^+ l^-\) Transition }
The total effective Hamiltonian, at the \(b-\)mass scale, for the quark-level transition \(b\rightarrow s l_i^+ l_j^-\) in the presence of NP operators is expressed as
\begin{equation}
\label{SM Hamiltonian}
\mathcal{H}_{eff}\left(b\rightarrow s l_i^+ l_j^-\right)=\mathcal{H}^{SM}_{eff}+\mathcal{H}^{NP}_{eff}=-\frac{4G_{F}}{\sqrt{2}}V^{*}_{ts}V_{tb} \sum_{p} C_{p}\mathcal O_{p},
\end{equation}
where
\begin{equation}
\label{total Hamiltonian}
\begin{aligned}
\mathcal{H}^{SM}_{eff}=-\frac{4G_{F}}{\sqrt{2}}V^{*}_{ts}V_{tb} \left\lbrace  \sum_{q=1}^{6} C_{q}(\mu)\mathcal O_{q}(\mu)+C_7\frac{e}{16\pi^2}\left[ \bar{s}\sigma_{\mu\nu}(m_sP_L+m_bP_R)b\right] F^{\mu\nu} \right. \\
 \left. +C_{9}^{ij}\frac{\alpha}{4\pi}\left( \bar{s}\gamma^{\mu}P_{L}b\right) \left( \bar{l_{i}}\gamma_{\mu}l_{j}\right) +C_{10}^{ij}\frac{\alpha}{4\pi}\left( \bar{s}\gamma^{\mu}P_{L}b\right) \left( \bar{l_{i}}\gamma_{\mu}\gamma_{5}l_{j}\right) \right\rbrace. 
\end{aligned}
\end{equation}
Here, \(P_{L,R}=\left( 1\mp\gamma_{5}\right) /2\) and \(\alpha=e^{2}/4\pi\) is the fine-structure constant. The six-dimensional operators \(\mathcal O_{q}\left( q=1,..6\right) \) correspond to \(P_{i}\) in Ref. \citen{Pi operators} and \(C_p\), (\(p=q,7,9,10\)) are the Wilson Coefficients. In the SM, only \(\mathcal{O}_{7} \), \(\mathcal{O}_{9} \) and \(\mathcal{O}_{10} \) are significant at the scale \(\mu=m_{b}\). As the analyses of the \(b\rightarrow s\) transitions indicate that the observed pattern of deviations is consistent with a larger Vector/Axial contribution \(\left( C^{\mu}_{9}, C^{\mu}_{10}\right) \), we will focus only on the \(V-A\) contributions that can only come from the neutral gauge bosons \(Z_{\mu}^{'}\),\(Z_{\mu}\), \(A_{\mu}\) and \(W_{\mu}^{6,7}\). The two dimension-six operators of interest are then
\begin{equation}
\begin{aligned}
\mathcal{O}_{9}^{ijkl}=\frac{\alpha}{4\pi}\left( \bar{d_k}\gamma^{\mu}P_{L}d_{l}\right) \left( \bar{l_{i}}\gamma_{\mu}l_{j}\right),\\
\mathcal{O}_{10}^{ijkl}=\frac{\alpha}{4\pi}\left( \bar{d_k}\gamma^{\mu}P_{L}d_{l}\right) \left( \bar{l_{i}}\gamma_{\mu}\gamma_5l_{j}\right),\\
\end{aligned}
\end{equation}
where \(i,j\) and \(k,l\) are the lepton and quark generation indices, respectively (\(k=2\) and \( l=3\) in our case).

\section{Gauge Bosons Contributions}
The expressions of the couplings of the charged and the neutral gauge bosons with the fermions are expressed in the interaction eigenbasis 
\subsection{Charged Currents}
The couplings of fermions to both the SM and the non-SM charged gauge bosons \(W_{\mu}^{\pm}\) and  \(Y_{\mu}^{\pm}\) respectively are given by the interaction Lagrangian density expressed in the interaction eigenbasis
\begin{equation*}
\label{lagCC}
\mathcal{L}_{CC} =\mathcal{L}_{W_{\mu}^{\pm}}+\mathcal{L}_{Y_{\mu}^{\pm}},
\end{equation*}
where
\begin{equation}
\label{lagW}
\begin{aligned}
\mathcal{L}_{W_{\mu}^{\pm}} = \dfrac{g}{\sqrt{2}} & W_{\mu}^{+} \left[(\overline{u_k}^L\gamma_{\mu}d_l^L) \delta_{k,l}+(\overline{\nu_i}^L\gamma_{\mu}e_j^L) \delta_{i,j}+\overline{N_4}^L\gamma_{\mu}E_4^L+\overline{(E_4^{-R})^c}\gamma_{\mu}N_5^{L}\right]\\
&+ W_{\mu}^{-} \left[(\overline{d_k}^L\gamma_{\mu}{u_l}^L) \delta_{k,l}+(\overline{e_i}^L\gamma_{\mu}{\nu_j}^L) \delta_{i,j}+\overline{E_4}^L\gamma_{\mu}{N_4}^L+\overline{N_5^{L}}\gamma_{\mu}(E_4^{-R})^c\right],
\end{aligned}
\end{equation}
and
\begin{equation}
\label{lagY}
\begin{aligned}
\mathcal{L}_{Y_{\mu}^{\pm}}= \dfrac{g}{\sqrt{2}} & Y_{\mu}^{+} \left[\overline{u_3}^L\gamma_{\mu}{D_3}^L+ \overline{U_m}^L\gamma_{\mu}{d_m}^L+\overline{{\nu}_n}^L\gamma_{\mu}{E_n}^L+\overline{N_1}^L\gamma_{\mu}{e_1}^L+ \overline{N_4}^L\gamma_{\mu}F_4^{-L}+\overline{({E_4^{-R}})^c}\gamma_{\mu}N_5^L\right]\\
&+Y_{\mu}^{-} \left[\overline{D_3}^L\gamma_{\mu}{u_3}^L+ \overline{d_m}^L\gamma_{\mu}{U_m}^L+\overline{{E_n}}^L\gamma_{\mu}{\nu}_n^L+\overline{e_1}^L\gamma_{\mu}{N_1}^L+ \overline{F_4^{-L}}\gamma_{\mu}N_4^L+\overline{P_5}^L\gamma_{\mu}(E_4^{-R})^c\right].
\end{aligned}
\end{equation} 
\(k\), \(l\) and \(i\), \(j\) refer to the SM quark and lepton generations, respectively.
\subsection{Neutral currents} 
The couplings of fermions to neutral gauge boson \(Z^{'}_{\mu}\), \(Z_{\mu}\), \(A_{\mu}\) and \(V^{\pm 0}_{\mu}\) (\(W_{\mu}^{6, 7}\)) are given by the interaction Lagrangian density expressed in the interaction eigenbasis
\begin{equation*}
\label{lagNC}
\mathcal{L}_{NC} =\mathcal{L}_{Z^{'}_{\mu}}+\mathcal{L}_{Z_{\mu}}+\mathcal{L}_{A_{\mu}}+\mathcal{L}_{V^{\pm0}_{\mu}},
\end{equation*}
where
\begin{equation}
\label{lagZ'}
\begin{aligned}
\mathcal{L}_{Z^{'}_\mu}=\dfrac{\cos\theta_{331}}{g_{x}}Z^{'}_{\mu} \left\lbrace\overline{U^L}\gamma_{\mu} \Gamma_{U}^{Z^{'}} U^L + \overline{D^L}\gamma_{\mu} \Gamma_{D}^{Z^{'}} D^L +\overline{f^L}\gamma_{\mu} \Gamma_{f^L}^{Z^{'}} f^L+ \overline{N^L}\gamma_{\mu} \Gamma_{N^L}^{Z^{'}} N^L \right.\\ 
\left. +\dfrac{g_{x}^2}{3\sqrt{6}}\left(2 \overline{U^R}\gamma_\mu U^R -\overline{D^R}\gamma_{\mu}D^R\right) +\overline{f^R}\gamma_{\mu}\Gamma_{f^R}^{Z^{'}}f^R\right\rbrace.
\end{aligned}
\end{equation}
\begin{equation}
\label{lagZ}
\begin{aligned}
\mathcal{L}_{Z_\mu}=\cos\theta_\omega g Z_{\mu} \left\lbrace \overline{U^L}\gamma_{\mu} \Gamma_{U}^Z U^L + \overline{D^L}\gamma_{\mu} \Gamma_{D}^Z D^{L} +\overline{f^L}\gamma_{\mu} \Gamma_{f^L}^Z f^{L}+\left( \dfrac{1+3\cos^2\theta_{331}}{2}\right) \overline{N^L}\gamma_{\mu}\Gamma_{N^L}^Z N^L \right.\\
\left. +\cos^2\theta_{331}\left( -2\overline{U^R}\gamma_\mu U_R+ \overline{D^R}\gamma_{\mu} D^{R}\right) +\overline{f^R}\gamma_{\mu}\Gamma_{f^R}^Z f^{R}\right\rbrace.
\end{aligned}
\end{equation}
\begin{equation}
\label{lagA}
\begin{aligned}
\mathcal{L_A}_{\mu}=\sqrt{3}\cos\theta_{331}\cos\theta_\omega g A_{\mu}\left\lbrace \dfrac{2}{3} \overline{U}\gamma_{\mu}U -\dfrac{1}{3} \overline{D}\gamma_{\mu}  D -\overline{f}\gamma_{\mu} f\right\rbrace. 
\end{aligned}
\end{equation}
\begin{equation}
\label{V}
\begin{aligned}
\mathcal{L}_{V^{\pm0}_{\mu}}=\dfrac{g}{\sqrt{2}}\left\lbrace V_0^{+} \left[ \overline{d_3^L}\gamma_\mu D_3^L - \overline{U_m^L}\gamma_\mu u_m^L - \overline{N_1^L}\gamma_\mu \nu_1^L + \overline{e^{-L}_n}\gamma_\mu E^{-L}_n + \overline{E^{-L}_4}\gamma_\mu F_4^{-L}+ \overline{N_5^{L}}\gamma_\mu P^L_5\right] \right. \\
\left. +V_0^{-} \left[  \overline{D^L_3}\gamma_\mu d^L_3 - \overline{u^L_m}\gamma_\mu U^L_m - \overline{\nu_1^L}\gamma_\mu N^L_1 + \overline{E^{-L}_n}\gamma_\mu e^{-L}_n+ \overline{F_4^{-L}}\gamma_\mu E^{-L}_4+\overline{P^L_5}\gamma_\mu N_5^{L}\right]\right\rbrace.
\end{aligned}
\end{equation}
The couplings to the neutral gauge bosons are proportional to the matrices
\begin{equation}
\Gamma_{U}^{Z^{'}}=diag\left( \frac{9g^2+g_x^2}{3\sqrt{6}},\frac{9g^2+g_x^2}{3\sqrt{6}},-\sqrt{\dfrac{3}{2}}g^2,\frac{-18g^2+g_x^2}{3\sqrt{6}},\frac{-18g^2+g_x^2}{3\sqrt{6}}\right).
\end{equation}
\begin{equation}
\Gamma_{D}^{Z^{'}}=diag\left( \frac{9g^2+g_x^2}{3\sqrt{6}},\frac{9g^2+g_x^2}{3\sqrt{6}},-\sqrt{\dfrac{3}{2}}g^2,\sqrt{6}g^2\right).
\end{equation}
\begin{equation}
\Gamma_{f^L}^{Z^{'}}=diag\left( \frac{9g^2-g_x^2}{3\sqrt{6}},\frac{-9g^2-2g_x^2}{3\sqrt{6}},\frac{-9g^2-2g_x^2}{3\sqrt{6}},\frac{18g^2-2g_x^2}{3\sqrt{6}},\frac{18g^2-2g_x^2}{3\sqrt{6}},\frac{-9g^2-g_x^2}{3\sqrt{6}},\frac{18g^2-2g_x^2}{3\sqrt{6}}\right).
\end{equation}
\begin{equation}
\Gamma_{N^L}^{Z^{'}}=diag\left(\frac{9g^2-g_x^2}{3\sqrt{6}},-\frac{9g^2+2g_x^2}{3\sqrt{6}},-\frac{9g^2+2g_x^2}{3\sqrt{6}},\frac{-18g^2-g_x^2}{3\sqrt{6}},-\frac{9g^2+2g_x^2}{3\sqrt{6}},\frac{-9g^2+g_x^2}{3\sqrt{6}},\frac{18g^2+g_x^2}{3\sqrt{6}}\right).
\end{equation}
\begin{equation}
\Gamma_{f^R}^{Z^{'}}=diag\left( -\frac{g_x^2}{\sqrt{6}},-\frac{g_x^2}{\sqrt{6}},-\frac{g_x^2}{\sqrt{6}},-\frac{g_x^2}{\sqrt{6}},-\frac{g_x^2}{\sqrt{6}},\frac{-9g^2+g_x^2}{3\sqrt{6}},-\frac{g_x^2}{\sqrt{6}}\right).
\end{equation}
\begin{equation}
\Gamma_{U}^Z=diag\left( \dfrac{1-\cos^{2}\theta_{331}}{2}, \dfrac{1-\cos^{2}\theta_{331}}{2}, \dfrac{1-\cos^{2}\theta_{331}}{2},-2\cos^{2}\theta_{331},-2\cos^{2}\theta_{331}\right).
\end{equation}
\begin{equation}
\Gamma_{D}^Z=diag \left(-\dfrac{1+\cos^{2}\theta_{331}}{2},-\dfrac{1+\cos^{2}\theta_{331}}{2},-\dfrac{1+\cos^{2}\theta_{331}}{2},\cos^{2}\theta_{331}\right). 
\end{equation}
\begin{equation}
\begin{aligned}
\Gamma_{f^L}^Z=diag \left(\dfrac{-1+3\cos^{2}\theta_{331}}{2},\dfrac{-1+3\cos^{2}\theta_{331}}{2},\dfrac{-1+3\cos^{2}\theta_{331}}{2},3\cos^{2}\theta_{331},3\cos^{2}\theta_{331}, \right. \\ \left.\dfrac{-1+3\cos^{2}\theta_{331}}{2}, 3\cos^{2}\theta_{331}\right). 
\end{aligned}
\end{equation}
\begin{equation}
\Gamma_{N^L}^Z=diag \left(1,1,1,0,1,-1,0\right).
\end{equation}
\begin{equation}
\Gamma_{f^R}^Z=diag \left(3\cos^{2}\theta_{331},3\cos^{2}\theta_{331},3\cos^{2}\theta_{331},3\cos^{2}\theta_{331},3\cos^{2}\theta_{331}, \dfrac{1-3\cos^{2}\theta_{331}}{2},3\cos^{2}\theta_{331}\right).   
\end{equation}
\section{The NP Contribution and Wilson Coefficients}
In what follows, we will consider only the contributions at the lowest order in \(\epsilon\) and we will focus only on the non-SM contributions to the Wilson Coefficients of interest \(C_{V_L}\), \(C_{9}^{\left( '\right) }\) and \( C_{10}^{\left( '\right) }\). We move to the mass eigenbases by introducing the unitary rotation matrices relating (unprimed) fermion interaction eigenstates and (primed) mass eigenstates.

\begin{equation}
\label{rotation}
\begin{aligned}
&q^{L}=V^{(q)}q^{'L}, &l^{L}=U^{(l)}l^{'L},& &f^{R}=W^{(f)}f^{'R},
\end{aligned}
\end{equation} 
where \(q\) and \(l\) stand for the quark and the lepton fields, respectively, while \(f\) stands for all fermion fields.
When diagonalizing the mass matrix using perturbation theory in powers of \(\epsilon\) we get \cite{main article}
\begin{itemize}
\item[(i)] at order \(\epsilon^{0}\), each of the massless SM particles and heavy exotic fermions mix only among themselves.
\item[(ii)] at order \(\epsilon^{1}\), there is only mixing between SM and exotic particles.
\item[(iii)] at order \(\epsilon^{2}\), there is mixing among all the particles of the same electric charge.
\end{itemize}
When considering only contributions at the lowest order in \(\epsilon\), we see that the \(b\rightarrow c\) transition that is mediated by the SM gauge bosons \(W_\mu^\pm\) occurs at \(\mathcal{O}(\epsilon^0)\) whereas the leading order non-SM contribution for leptons arises at \(\mathcal{O}(\epsilon^2)\). For the heavy gauge boson, on the other hand, the quark transition does not occur at \(\mathcal{O}(\epsilon^{0})\). It arises, however, at \(\mathcal{O}(\epsilon^{1})\), where \(Y_{\mu}^{+}\) always couples an SM particle with an exotic one. In addition to that, the Hamiltonian contains a \(\mathcal{O}(\epsilon^{2})\) suppression (compared to the SM) that comes from the heavy mass in the propagator of the gauge boson. Therefore, the \(Y_{\mu}^{+}\) contribution starts at an already high overall \(\mathcal{O}(\epsilon^{3})\) order which can be neglected compared to the \(\mathcal{O}(\epsilon^{2})\) contribution from the \(W_{\mu}^{+}\) gauge boson regardless of what the order of the leading contribution of the leptons might be. As for the neutral transition  \(b\rightarrow s\), it is clear that the restriction of the interaction matrix \(\Gamma_{D}^{Z^{'}}\) to the SM particles is not proportional to the identity matrix in flavor space. So, the FCNC transition arises already at \(\mathcal{O}(\epsilon^{0})\), and due to the \(\mathcal{O}(\epsilon^{2})\) suppression, compared to the SM, that results from the heavy mass of the \(Z_{\mu}^{'}\) boson, we conclude that the NP contribution from the \(Z_{\mu}^{'}\) boson starts at \(\mathcal{O}(\epsilon^{2})\). Thus, we consider only \(\mathcal{O}(\epsilon^{0})\) SM terms of the lepton sector. 
In the case of the neutral SM gauge boson, however, the \(b\rightarrow s\) transition arises at \(\mathcal{O}(\epsilon^{2})\), and because there is no \(\mathcal{O}(\epsilon^{2})\) suppression due to the gauge boson mass, the NP contribution for the \({Z_{\mu}}\) starts also at \(\mathcal{O}(\epsilon^{2})\). Thus, as in the \(Z_{\mu}^{'}\), we consider only \(\mathcal{O}(\epsilon^{0})\) SM terms of the lepton sector. The interactions of the right handed quarks, with both neutral bosons, are proportional to the identity in flavor space (Eqs. \eqref{lagZ'}, \eqref{lagZ}), so no flavor change can arise at any order in \(\epsilon\). We conclude that \(Z_{\mu}^{'}\) and \(Z_{\mu}\) do not contribute to \(C_{9}^{'}\) and \(C_{10}^{'}\). Our model does not allow for any contribution to \(C^{'}_{9,10}\) in the process.\\
For the photon \(A_{\mu}\), the interaction with down-type quarks is proportional to the identity matrix in the flavor space \eqref{lagA}, so there are no FCNC from the photon interaction. As for the heavy \(W_{V^{\pm0}}\), the contributions \eqref{V} to the process is also of the \(\mathcal{O}(\epsilon^{3})\). It can be neglected compared to \(Z_{\mu}^{'}\) and \(Z_{\mu}\)'s.
\subsection{Contribution from the Charged Gauge Bosons (\(C_{VL}\))}
The (leading order) \(\mathcal{O}(\epsilon^2)\) effective Hamiltonian for the flavor changing charged transition mediated by the \(W_\mu^+\) is thus 
\begin{equation}
\begin{aligned}
\label{W Hamiltonian}
\mathcal{H}^{W^+}_{eff} \supset \frac{g^2}{2 M_W^2}\sum_{n=1,2,3}(V^{*(u)}_{nk}V^{(d)}_{nl})\left[  \sum_{n=1,2,3}(U^{*(\nu)}_{ni}U^{(e)}_{nj})+U^{*(\nu)}_{4i}U^{(e)}_{4j}+ U^{*(\nu)}_{5i}W^{(e)}_{5j}\right] (\bar{u}_k^L \gamma_\mu d_l^L)(\bar{\nu}_i^L \gamma^\mu e_j^L),
\end{aligned}
\end{equation}
where \(n=1, 2, 3\) are SM mass indices, \(k, l\) and \(i, j\) are quark and lepton flavor indices, respectively. The superscripts (\(\nu\)) and (\(e\)) stand for the neutral and the charged lepton, respectively. For the charged transition of interest (\(k=3\) and \(l=2\)), \(\sum_{n=1,2,3}(V^{*(u)}_{nk}V^{(d)}_{nl})\) is nothing but the \(V_{CKM}\) element \(V_{cb}\) (Eq. \eqref{vckm}) and \(\sum_{n=1,2,3}(U^{*(\nu)}_{ni}U^{(e)}_{nj})\) is the element \((U_{PMNS})^{SM}_{ij}\) (see \ref{appB}). Thus, exploiting the unitarity of the SM's \(U_{PMNS}\) matrix at the order of interest for \(i=j\)
\begin{equation}
\label{W final Hamiltonian}
\begin{aligned}
\mathcal{H}^{W^+}_{eff} \supset \frac{g^2}{2 M_W^2}V_{cb}\left[ 1+U^{*(\nu)}_{4i}U^{(e)}_{4i}+ U^{*(\nu)}_{5i}W^{(e)}_{5i}\right] (\bar{c}\gamma_\mu b)(\bar{\nu}_i \gamma^\mu e_i).
\end{aligned}
\end{equation}
Comparing to equation \eqref{SM Hamiltonian}, the model's contribution to the operator \(\mathcal{O}_{V_{L}}^{ll}\) is thus  \(C_{V_L} = U^{*(\nu)}_{4i}U^{(e)}_{4i}+ U^{*(\nu)}_{5i}W^{(e)}_{5i}\).
Both (\(7\times7\)) rotation matrices \(U^{\nu}\) and \(U^e\) are unitary, and so is their product, so we can write for a specific SM generation \(I\)
\begin{equation}
\label{LambdaL}
\begin{aligned}
U^{*(\nu)}_{4I}U^{(e)}_{4I}=1-\sum_{(J\neq I)J=1}^{7}U^{*(\nu)}_{4J}U^{(e)}_{4J}.
\end{aligned}
\end{equation}
The same can be said about the product (\(7\times7\)) of the two unitary matrices \(U^{(\nu)}\) and \(W^{(e)}\) 
\begin{equation}
\label{LambdaR}
\begin{aligned}
U^{*(\nu)}_{5I}W^{(e)}_{5I}= 1-\sum_{(J\neq I)J=1}^{7}U^{*(\nu)}_{5J}W^{(e)}_{5J}.
\end{aligned}
\end{equation}
\(J\) represents all the remaining generations. We denote with \(\Lambda_I=U^{*(\nu)}_{4I}U^{(e)}_{4I}\) and \(\Delta_I=U^{*(\nu)}_{5I}W^{(e)}_{5I}\). Equations \eqref{LambdaL} and \eqref{LambdaR} mean that \(\Lambda_I\) and \(\Delta_I\) should belong each to \([0,1[\), so \(\Lambda_I +\Delta_I\) should also remain within the interval \([0,1[\) in order for the term to account as a contribution which agrees with the allowed \(1\sigma\) range for the effective coefficient \(C_{V_L}\) \cite{Cvl} for \(\Lambda_I +\Delta_I\in [0.09, 0.13]\). We see that our model provides a good explanation of the Vector/Axial LFUV contribution to \(b\rightarrow cl\bar{\nu}_{l}\) transitions provided that the dominant contributions come from the gauge bosons rather than the Higgs sector. 

\subsection{Contributions from the Neutral Gauge Bosons (\(C_9, C_{10}\))}
Moving to the mass eigenbasis, and exploiting the unitarity of the rotation matrices \(V\), \(U\) and \(W\) for the \(\delta_{ij}\) contribution, we eliminate the coupling \(g\) by means of Eq. \eqref{relation between two gauge coupling constants}. The leading-order \(Z_{\mu}^{'}\) and \(Z_{\mu}\) contributions in terms of effective operators read (\(V_{(3,4)k}^{*}V_{(3,4)l} \equiv V_{(3,4)k}^{(d)*}V^{(d)}_{(3,4)l}\) is understood, where (\(d\)) stands for a down-type quark, and \(U_{1i}^{*}U_{1j} \equiv U_{1i}^{(e)*}U^{(e)}_{1j}\) where the superscript (\(e\)) stands for a charged lepton)
\begin{equation}
\begin{aligned}
\mathcal{H}^{Z^{'}}_{eff}\supset\frac{g_{x}^{2}}{108\cos^{2}\theta_{331}}\frac{1}{M_{Z^{'}}^{2}}V_{3k}^{*}V_{3l}\frac{4\pi}{\alpha}
\left\lbrace \left[ \left( \frac{1+9\cos^{2}\theta_{331}}{2}\right)  \delta_{ij}-U_{1i}^{*}U_{1j}\right] \mathcal{O}_{9}^{ijkl} \right.\\
\left. +\left[ \left( \frac{3\cos^{2}\theta_{331}-1}{2}\right) \delta_{ij}+U_{1i}^{*}U_{1j}\right]  \mathcal{O}_{10}^{ijkl}\right\rbrace,
\end{aligned}
\end{equation}
and
\begin{equation}
\begin{aligned}
&\mathcal{H}^{Z}_{eff}\supset\frac{\cos^{2}\theta_{W}\left( 1+3\cos^{2}\theta_{331}\right) }{8}\frac{g^{2}}{M_{Z}^{2}}\frac{4\pi}{\alpha} V_{4k}^{*}V_{4l} \delta_{ij} \left[ \left( -1+9\cos^{2}\theta_{331}\right) \mathcal{O}_{9}^{ijkl}+\left( 1+3\cos^{2}\theta_{331}\right) \mathcal{O}_{10}^{ijkl}\right]. &
\end{aligned}
\end{equation}
Keeping the same notation as in Ref.\citen{main article}, the NP contributions from \(Z_{\mu}\) and \(Z_{\mu}^{'}\) to the Wilson Coefficients can be written in terms of the quantities \(f_{Z^{'}}\) and \(f_{Z}\) as
\begin{equation}
C_{9}^{ij}=f_{Z^{'}}\left( -\lambda_{ij}+\frac{1+3\tan^{2}\theta_{W}}{2}\delta_{ij}\right)+f_{Z}\left( -1+3\tan^{2}\theta_{W}\right) \delta_{ij},
\end{equation}
and
\begin{equation}
C_{10}^{ij}=f_{Z^{'}}\left( \lambda_{ij}+\dfrac{\tan^{2}\theta_{W}-1}{2}\delta_{ij}\right)+f_{Z}\left( 1+\tan^{2}\theta_{W}\right) \delta_{ij},
\end{equation}
where
\begin{equation}
\label{fz'}
\begin{aligned}
&f_{Z^{'}}=-\frac{1}{2\sqrt{2}G_{F}V_{tb}V^{*}_{ts}}\frac{4\pi}{\alpha}\frac{1}{6-2\tan^{2}\theta_{W}}\frac{g^{2}}{M_{Z^{'}}^{2}}V^{*}_{3k}V_{3l},&
\end{aligned} 
\end{equation}
and
\begin{equation}
\label{fz}
\begin{aligned}
&f_{Z}=-\frac{1}{2\sqrt{2}G_{F}V_{tb}V^{*}_{ts}}\frac{4\pi}{\alpha}\frac{1}{8}\frac{g^{2}}{M_{Z}^{2}}V^{*}_{4k}V_{4l}.&
\end{aligned}
\end{equation}
\(\lambda_{ij}=U^{*}_{1i}U_{1j}\). Even though our model allows for lepton flavor violating transition with different leptons in the final state \((i\neq j)\), these processes have not been observed yet, so, assuming that they are suppressed, we set their coefficients to zero. The solution \(f_{Z^{'}}=0\), i. e. the NP contribution is zero, should be discarded as it would mean the absence of LFUV. So, we are left with \(\lambda_{ij}=0\) for \(i\neq j\).
By definition\\
\begin{equation}
\label{lambdaij}
\lambda_{ij}=0\Longrightarrow U^{*}_{1i}U_{1j}=0.
\end{equation}
Equation \eqref{lambdaij} does not necessarily imply that both \(U\) matrix elements have to be zero; one rotation matrix entry can be non-zero for a generation \(i\) (e.g. \(i=1\)) while the other two entries (e.g. \(j=2,3\)) are zero, ensuring that the above annihilation is realized. We denote with \(I\) the generation for which the entry for the rotation matrix is non-zero, and with \(i\) the other generations. We get
\begin{equation}
\label{CI}
\begin{aligned}
C_{9}^{I}=f_{Z^{'}}\left( -\lambda_{I}+\frac{1+3\tan^{2}\theta_{W}}{2}\right) +f_{Z}\left( -1+3\tan^{2}\theta_{W}\right),\\
C_{10}^{I}=f_{Z^{'}}\left( \lambda_{I}+\frac{\tan^{2}\theta_{W}-1}{2}\right) +f_{Z}\left( 1+\tan^{2}\theta_{W}\right),
\end{aligned}
\end{equation}
and
\begin{equation}
\label{Ci}
\begin{aligned}
C_{9}^{i}=f_{Z^{'}}\left(\frac{1+3\tan^{2}\theta_{W}}{2}\right) +f_{Z}\left( -1+3\tan^{2}\theta_{W}\right),\\
C_{10}^{i}=f_{Z^{'}}\left( \frac{\tan^{2}\theta_{W}-1}{2}\right) +f_{Z}\left( 1+\tan^{2}\theta_{W}\right).
\end{aligned}
\end{equation}
Inverting relations \eqref{Ci} we get
\begin{equation}
\label{Z' contribution}
f_{Z^{'}}=\frac{1+\tan^{2}\theta_{W}}{4\tan^{2}\theta_{W}}C_{9}^{i}-\frac{-1+3\tan^{2}\theta_{W}}{4\tan^{2}\theta_{W}}C_{10}^{i}.
\end{equation}
From the system of equations \eqref{Ci} and \eqref{CI} we get
\begin{equation}
\label{relation between Ci, CI and Z' contribution}
2\lambda_{I}f_{Z^{'}}=C_{10}^{I}-C_{9}^{I}-C_{10}^{i}+C_{9}^{i}.
\end{equation}
We now have to identify which index corresponds to which lepton, knowing that, based on phenomenological constraints, the electronic NP contribution to the effective Hamiltonian \(C_{9,10}^e\) is absent. 
\begin{itemize}
\item[(i)]
If we identify the electron with the index \(i\) (for which the entry for the rotation matrix vanishes), we set \(C_{9,10}^{i}=0\). Equation \eqref{Z' contribution} implies that \(f_{Z^{'}}=0\), solution that has to be discarded since it would mean no LFUV.
\item[(ii)] If the electron is identified with the index \(I\), the coefficients \(C_{9,10}^{I}\) must be set to zero in this case, and the remaining index \(i\) would correspond to the other two generations. In this case, eq. \eqref{relation between Ci, CI and Z' contribution} yields constraints on the non-vanishing NP Wilson Coefficients for \(\mu\) and \(\tau\)
\begin{equation}
\label{final result}
\frac{C_{9}^{\mu}}{C_{10}^{\mu}}=\frac{2\tan^{2}\theta_{W}+\lambda_{e}\left( 1-3\tan^{2}\theta_{W}\right) }{2\tan^{2}\theta_{W}-\lambda_{e}\left( 1+\tan^{2}\theta_{W}\right) }.
\end{equation}
\end{itemize}
Due to the unitarity of the (\(7\times7\)) rotation matrix \(U\), we have
\begin{equation}
\lambda_{I}=\mid U_{1I} \mid^{2}=1-\sum_{(i\neq I)i=1}^{7} \mid U_{1i} \mid^{2},
\end{equation}
which means that \(0<\lambda_{e}\leq 1\). Imposing that \(\mid C^{\mu}_{10}\mid\leq\mid C^{\mu}_{9}\mid\), the one-dimensional scenario of the global analysis that favors NP in \(C^{\mu}_{9}=-C^{\mu}_{10}\) within the \(1\sigma\) interval \([-0.75, -0.49]\) \cite{global analysis} is explained for \(0.71\leq\lambda_e\leq0.86\) (\(\theta_W\simeq 29^{\circ}\)). The exact equality \(C_{9}^{\mu}/C_{10}^{\mu}=-1\), obtained for \(\lambda_{e}=1\), which means that the left-handed interaction of the electron is a mass eigenstate, is also allowed for this case. It is worth mentioning that in case A of the set with \(\beta=-1/\sqrt{3}\)\cite{main article}, the allowed region to the Wilson Coefficients imposing that \(\mid C^{\mu}_{9}\mid\leq\mid C^{\mu}_{10}\mid\) is explained for \(0.81\leq\lambda^L_e\leq1\), which agrees with our set, where \(\mid C^{\mu}_{10}\mid\) is supposed to be less than \(\mid C^{\mu}_{9}\mid\), only for  \(\lambda_{e}=1\) \cite{mine}. The case B in Ref. \citen{main article}, however, is not taken into account within our model since the right-handed components are not concerned with the modification applied to the fields. The other two scenarios, however, (NP in \(C_{9}^{\mu}=-C_{9{'}}^{\mu}\) and NP in \(C_{9}^{\mu}\) \cite{global analysis}) cannot be described in the framework of our model. Since no FCNC arises for right-handed quarks due to their diagonal interaction terms in flavor space (Eqs. \eqref{lagZ} and \eqref{lagZ'}), \(C_{9'}^{\mu}=0\). 
Figure \eqref{Fig.1} shows the allowed region for the Wilson Coefficients as indicated by the gray wedge imposing \(C_{9}^{\mu}/C_{10}^{\mu}\) to remain between \(-1\) and \(-2.04\). The light grey wedge are the results obtained for \(0.71\leq\lambda_e\leq1\). The dark gray wedge represents the the one-dimensional scenario of the global analysis that favors NP in \(C^{\mu}_{9}=-C^{\mu}_{10}\) within the \(1\sigma\) interval \([-0.75, -0.49]\).\\
\begin{figure}
\centerline{\includegraphics[width=7cm]{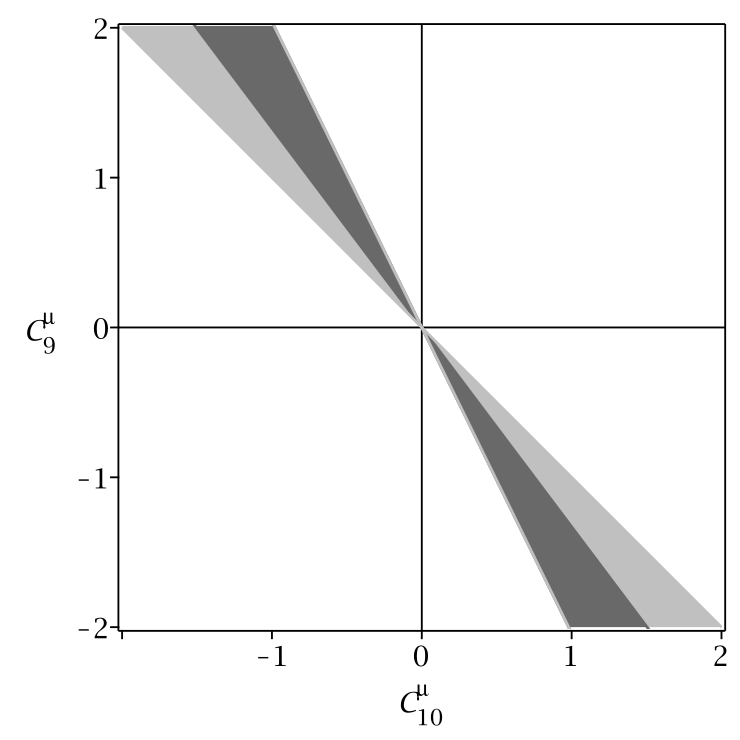}}
\caption{Allowed regions for the Wilson Coefficients for \(\beta=1/\sqrt{3}\) imposing that \(\mid C^{\mu}_{10}\mid\leq\mid C^{\mu}_{9}\mid\). The dark gray wedge shows the favored one-dimensional NP scenario in \(C^{\mu}_{9}=-C^{\mu}_{10}\).  \label{Fig.1}}
\end{figure}
In summary, the electron (first generation of SM leptons) has to be identified with the non-vanishing entry in the rotation matrix \(U\) in order to have non-vanishing NP contributions to Wilson Coefficients, for both \(\mu\) and \(\tau\) that agree with the favored one-dimensional scenario of NP in \(C_{9}^{\mu}=-C_{10}^{\mu}\).
\section{Conclusion}
In an attempt to give an explanation to the deviations from the Standard Model in both charged and neutral flavor changing transitions \(b\rightarrow cl\bar{\nu}_{l}\) and \(b\rightarrow s l^+ l^-\), respectively, we have investigated a non-minimal version of the 331 models. In order to explain LFUV, five lepton triplets are required in this set, where the additional heavy gauge bosons and fermions have electric charges similar to those of the SM particles. Some modifications had to be applied to the charged lepton sector in order to have a physically allowed particle spectrum at the end of the two SSBs. We have shown how this model could explain these experimentally observed deviations, provided that the latter are dominated by the contributions from the charged and neutral gauge bosons \(W^{\pm}_{\mu}\), \(Y^{\pm}_{\mu}\) and \(Z_{\mu}\), \(Z^{'}_{\mu}\), respectively. For the charged transition, the contribution to the dimension 6 operator \((\bar{c}\gamma_\mu b)(\bar{\nu}_i \gamma^\mu e_i)\) could be explained. In fact, the effective NP coupling \(C_{VL}\) derived within the framework of our model agrees with the \(1\sigma\) solution favored by the global fits. It is shown that the contribution arises at \(\mathcal{O}(\epsilon^2)\) through the couplings induced by mixings of the light charged gauge boson with the fermions, but not of the exotic gauge boson since they are suppressed by its heavy mass. As for the neutral transition \(b\rightarrow s l^+ l^-\), our model shows no ability to accommodate NP contribution in \(C_{9}^{\mu}=-C_{9^{'}}^{\mu}\) due to the absence of the FCNC for the right-handed quarks. It is shown that, as in the charged current case, the neutral contributions  arise at \(\mathcal{O}(\epsilon^2)\) for both neutral bosons (\(Z_{\mu}\), \(Z^{'}_{\mu}\)) mediated transitions. When constraints on the mixing matrices between interaction and mass fermion eigenstates are put in light of the absence of contributions to \(b\rightarrow s e^+ e^-\), our model becomes able to accommodate significant NP contribution in \(C_{9}^{\mu}=-C_{10}^{\mu}\), in agreement with an NP scenario favored by global fits. 
It is worth stressing that, despite being left out of the discussion, neutrino spectrum requires an accurate analysis within our model. It has been shown that neutrino masses could be generated in the 331 models at the \(TeV\) scale either through seesaw mechanism \cite{seesaw, seesaw2}, or, by exploiting the tree level realization of dimension-five effective operators \cite{D5}. Nonetheless, exploring the neutrino mass spectrum is of great importance.

\section*{Acknowledgments}

We are very grateful to the Algerien Ministry of Higher Education and Scientific Research and the DGRSDT for the finantial support.
\appendix
\section{Yukawa Terms}
\label{appA}
The model undergoes two stages of Spontaneous Symmetry Breakdown (SSB). The first SSB (\(331\rightarrow 321\)) occurs at the \(TeV\) scale by means of the \(vev\)s of both a triplet \(\chi\) and a sextet \(S\) \cite{scalar sector}
\begin{equation}
\begin{aligned}
\label{first SSB}
<\chi>=\frac{1}{\sqrt{2}}
\begin{pmatrix}
0\\0\\u
\end{pmatrix}
\sim (1, 3, \frac{1}{3}) &
;& <S>=
\begin{pmatrix}
0& 0&  0\\
0& 0& 0\\
0& 0& a\\
\end{pmatrix}
\sim (1, 6, \frac{2}{3}).
\end{aligned}
\end{equation}
At this stage, the Yukawa Lagrangian containing all gauge invariant lepton-lepton-scalar terms for the negatively charged leptons reads
\begin{equation}
\begin{aligned}
\label{Yukawa Lagrangian}
\mathcal{L}_Y^{l^-}\supset \chi^* \bar{l}^L_{n,4} (l_{5}^L)^c + \chi \bar{l}^L_{n,4} \left( e^{-R}_m + E^{-R}_{n,4} + E^{-R}_5\right).
\end{aligned}
\end{equation}
Where the combination of \(SU(3)_{L}\) triplets and antitriplets is
\begin{equation}
\epsilon_{ijk}(\chi^{*i} )\bar{l}^{Lj}_{n,4} (l_{5}^L)^{ck}.
\end{equation}
whereas the Yukawa terms that can be built with the sextet lead to Majoranna masses for the exotic neutral leptons \(N_5\) and \(N_1\). They are of the form \(S\bar{l}_5 (l_5^c+ l_1)\).
Originally, the model consisted of 5 exotic charged leptons where one positively charged lepton belongs to the fifth generation
\begin{equation*}
\begin{aligned}
&l_{5}^{L}=
\begin{pmatrix}
E^{+L}_{5}\\P^{L}_{5}\\N^{L}_{5}
\end{pmatrix}\sim \left( 1, 3, 1/3 \right),&
E_{5}^{+R} \sim \left( 1, 1, 1 \right).
\end{aligned}
\end{equation*}
The (\(TeV\) scale) Yukawa Lagrangian for the negatively charged leptons reads
\begin{equation}
\begin{aligned}
\label{Yukawa Lagrangian for charged leptons}
\mathcal{L}_Y^{l^-}\supset&-\dfrac{u^*}{\sqrt{2}}\left( \sum_{n=2,3}Y_{n,5}\bar{e}^{-L}_nE_5^{-L}+Y_{4,5}\bar{E}_4^{-L}E_5^{-L}\right)\\&+\dfrac{u}{\sqrt{2}}\sum_{n=2,3}[\left(  \bar{E}_n^{-L}+F_4^{-L}\right) (y^{-}_{n,1} e_1^{-R}+y^{-}_{n,2} e_2^{-R}+ y^{-}_{n,3}e_3^{-R}+y^{'-}_{n,2}E_2^{-R}+y^{'-}_{n,3}E_3^{-R}\\&+y^{-}_{n,4}E_4^{-R}+y^{'-}_{n,4}F_4^{-R}+y^{-}_{n,5}E_5^{-R})]. 
\end{aligned}
\end{equation}
Where  \(y^{(')}_n\) and \(Y_n\) are the Yukawa couplings. At the this stage, \(E_{5}^{-}\) is masseless (together with \(E_{4}^{-}\)). Meaning that, at the second (EW scale) SSB, not only the 3 SM charged leptons would acquire mass, but also the two exotic ones, and because the spectrum should contain no light particles apart from the SM ones, we have to get rid of such presence. To do so, we identify the \(E^{+L}_{5}\) with the charge conjugate of the right handed component of \(E^{-}_{4}\). Thus, the right handed part of \(E^{-}_{4}\) should belong to the lepton triplet \(l_{5}^{cL}\) (\((E^{-R}_{4})^c \sim (1, 3, \frac{1}{3})\)) rather than being a singlet. The second SSB is triggered by three sextets and two triplets. Their \(vev\)s are aligned as follows
\begin{equation}
\begin{aligned}
\label{second SSB}
<\eta>=\frac{1}{\sqrt{2}}
\begin{pmatrix}
0\\w_2\\w_3
\end{pmatrix}
\sim (1, 3, \frac{1}{3}); 
&<\rho>=\frac{1}{\sqrt{2}}
\begin{pmatrix}
v\\0\\0
\end{pmatrix}
\sim (1, 3, -\frac{2}{3}); &\\
<S_b>=
\begin{pmatrix}
b& 0&  0\\
0& 0& 0\\
0& 0& 0\\
\end{pmatrix}
\sim (1, 6, -\frac{4}{3});&
<S_c>=
\begin{pmatrix}
0& c_1&  c_2\\
c_1& 0& 0\\
c_2& 0& 0\\
\end{pmatrix}
\sim (1, 6, -\frac{1}{3});&\\
<S_d>=
\begin{pmatrix}
0& 0&  0\\
0& d_1& d_2\\
0& d_2& d_3\\
\end{pmatrix}
\sim (1, 6, \frac{2}{3}).&
\end{aligned}
\end{equation} 
It should be emphasized here on the fact that this is the only possible set for \(\beta=1/\sqrt{3}\). Any other attempt to limit the number of the extra degrees of freedom would lead to Yukawa terms that, despite being allowed by symmetry arguments, are forbidden by mass scale arguments (e. g. identifying \(F_4^{-L}\) with \(e^{-R}_{1}\) would lead to a \(TeV\) Yukawa term \(\propto e_1^{-R}e_2^{-R}\)). After the two SSBs, the most general Yukawa Lagrangian responsible for the charged lepton masses reads
\begin{equation}
\label{Charged lepton Yukawa lagrangian}
\begin{aligned}
\mathcal{L}_Y^{l^-}=&\left[\bar{l}_n^L\left(y_{n}\chi+k_{n}\eta\right)+\bar{l}_1^L h_{1}\rho^*\right]l^R
+\bar{l}^L_n l_5^{cL}\left(Y_{n5}\chi^* + K_{n5}\eta^* + C_{n5}S_c\right),
\end{aligned}
\end{equation}
where
\begin{itemize}
\item[•] \(l^R = \left( e^{-R}_{1},e^{-R}_{2},e^{-R}_{3},E^{-R}_{2},E^{-R}_{3},F^{-R}_{4}\right)\).
\item[•] \(y_n\), \(k_n\) and \(h_n\) are the Yukawa couplings of the left handed \(n\) and the right handed charged lepton fields with the scalars \(\chi\), \(\eta\) and \(\rho\), respectively.
\item[•] \(Y_{nn}\) and \(K_{nn}\) are the Yukawa couplings of the left handed lepton fields \(n\) with the scalars \(\chi\) and \(\eta\) respectively.
\item[•] \(C_{nn}\) are  the Yukawa couplings of the left handed lepton fields \(n\) with the sextet \(S_c\).
\end{itemize}  
The spectrum thus contains three EW mass (SM) plus four \(TeV\) mass exotic charged leptons.\\
The Yukawa Lagrangian responsible for the quark masses is:
\begin{equation}
\label{Quark Yukawa lagrangian}
\begin{aligned}
\mathcal{L}_Y^{q}=&\left(\bar{Q}_L^m \chi^*Y_{mi}^u+\bar{Q}_L^m \eta^*j_{mi}^u+\bar{Q}_L^3 \rho y_{3i}^u\right)U_i^R+\left(\bar{Q}_L^3 \chi Y_{3j}^d+\bar{Q}_L^3 \eta j_{3j}^d+\bar{Q}_L^m \rho^* y_{mj}^d\right)D_j^R,
\end{aligned}
\end{equation}
where
\begin{itemize}
\item[•] \(U_i^R = \left( u_1^R,u_2^R,u_3^R, U_1^R,U_2^R\right)\).
\item[•] \(D_j^R = \left( d_1^R,d_2^R,d_3^R, D_3^R\right)\).
\item[•] \(Y^{u,d}\), \(y^{u,d}\) and \(j^{u,d}\) are the Yukawa couplings for \(\chi\), \(\rho\) and \(\eta\), respectively.
\end{itemize}  
\section{\(V_{CKM}\) and \(U_{PMNS}\) matrices in the 331 model}
Despite being unitary in the SM, both \(V_{CKM}\) and \(U_{PMNS}\) are not unitary in the 331 model due to the presence of \(\varphi\) and \(\xi\) matrices, respectively. Recovering unitary  \(V_{CKM}\) and \(U_{PMNS}\) matrices requires remaining at low energies (i. e. leading order in \(\epsilon\)) by considering only their \(3\times 3\) SM block matrices. 
\label{appB}
\subsection{\(V_{CKM}^{331}\) Matrix}
The Cabibbo-Kobayashi-Maskawa (CKM) matrix is given by the \(W_{\mu}^+\) interaction with quarks
\begin{equation}
\begin{aligned}
\frac{g}{\sqrt{2}}W_{\mu}^+ \bar{U}^L\gamma_\mu \varphi D^{L} = \frac{g}{\sqrt{2}}W_{\mu}^+ \bar{U}^L\gamma_\mu
\begin{pmatrix}
1 & 0 & 0 & 0 \\
0 & 1& 0 & 0  \\
0 & 0 & 1 & 0 \\
0 & 0 & 0 & 0\\
0 & 0 & 0 & 0\\
\end{pmatrix}
D^{L},
\end{aligned}
\end{equation}
with the \(5\times 4\) matrix \(V_{CKM}^{331}=V^{(u)^\dagger} \varphi V^{(d)}\) which elements can be written as 
\begin{equation}
\begin{aligned}
\label{vckm}
{V_{CKM}^{331}}_{(k,l)}=\sum_{n=1,2,3}V^{*(u)}_{nk}V^{(d)}_{nl},
\end{aligned}
\end{equation}
where \(n=1,2,3\) are the SM mass indices and \(k=1,..,5\) and \(l=1,..,4\) are flavor indices.
\subsection{\(U_{PMNS}^{331}\) Matrix}
The Pontecorvo-Maki-Nakagawa-Sakata (PMNS) matrix is built by combining the two unitary rotation matrices \(U^{(e)}\) and \(U^{(\nu)}\) as 
\begin{equation}
\begin{aligned}
\dfrac{g}{\sqrt{2}}W_{\mu}^{+} \bar{N}^{L}\gamma_{\mu} \xi f^{-L}=\dfrac{g}{\sqrt{2}}W_{\mu}^{+} \bar{N}^L\gamma_{\mu}
\begin{pmatrix}
1& 0 & 0 & 0& 0 & 0  \\
0 & 1 & 0 & 0 & 0 & 0  \\
0 & 0 & 1 & 0 & 0 & 0  \\
0 & 0 & 0 & 1  & 0 & 0  \\
0 & 0 & 0 & 0 &  0  & 0  \\
0 & 0 & 0 & 0 & 0 &0  \\
\end{pmatrix}
f^{-L},
\end{aligned}
\end{equation}
where 
\begin{equation*}
\begin{aligned}
&f^{-L}=\left( e_1^{-L}, e_2^{-L}, e_3^{-L}, E_4^{-L}, F_4^{-L}, E_2^{-L}, E_3^{-L}\right)^{T}. &\\ 
&N^L=\left( \nu_1^L,\nu_2^L,\nu_3^L, N_4^L, N_5^L, P_5^L, N_1^L \right)^{T}. &
\end{aligned}
\end{equation*}
Its elements can be expressed as 
\begin{equation}
\left( U_{PMNS}^{331}\right)_{ij} =\left( U_{PMNS}^{SM}\right)_{ij} +U^{*(\nu)}_{4i}U^{(e)}_{4j},
\end{equation} 
where \(i,j\) are lepton generation indices. The \(U_{PMNS}^{331}\) could be developped for a non-minimal 331 model with \(\beta=1/\sqrt{3}\) by obtaining bounds on the mixing angles in the leptonic sector studying LFV processes \cite{UPMNS}, but given the current study, exploiting the unitarity of the rotation matrices \(U^{(\nu)}\) and \(U^{(e)}\) has proved sufficient for the problem at hand.

\end{document}